\documentclass[letter]{aa}
\usepackage{amssymb,amsmath}
\usepackage{graphicx}
\usepackage{xcolor,natbib}
\usepackage{multirow}
\usepackage[T1]{fontenc}
\usepackage{ae,aecompl}
\usepackage{newtxtext, newtxmath}
\usepackage{float}
\usepackage{subfigure}
\usepackage{longtable}
\usepackage{enumitem}
\usepackage{longtable,listings}
\usepackage[flushleft]{threeparttable}
\usepackage{parcolumns}
\usepackage{hyperref}
\usepackage{lineno}

\def\obj{4C 50.43}

\begin{document}

\title{Optical QPOs with different periodicities in CSS and ZTF light curves of the quasar 4C 50.43}	

\titlerunning{Different Periodicities in different periods}

\author{GuiLin Liao\inst{1}, XingQian Chen\inst{1}, PeiZhen Cheng\inst{1}, 
Xue-Guang Zhang\inst{1*}}
	
\institute{Guangxi Key Laboratory for Relativistic Astrophysics, School of Physical Science and Technology, GuangXi
University, No. 100, Daxue Road, Nanning, 530004, P. R. China \ \ \ \email{xgzhang@gxu.edu.cn}}

\date{}

\abstract{
	Long-standing optical quasi-periodic oscillations (QPOs) with periodicity of hundreds to thousands of days have been 
accepted as indicators for central sub-pc binary black hole systems (BBHs) in broad line active galactic nuclei (BLAGN). 
However, there are so far no direct reports on whether such reported optical QPOs have their periodicities constant in 
different periods. Here, based on different methods applied to light curves of 4C 50.43 in different periods, optical QPOs 
with periodicity of 1124days was detected in the CSS V-band light curve, while a shorter periodicity of 513days was detected 
in the ZTF g/r band light curves. Despite the two periodicities near-harmonic 2:1 ratio, their absence of simultaneous 
detection in the lomb-scargle periodograms of the ZTF light curves suggests that they are unlikely to be harmonically related. 
Potential factors were considered to explain these two distinct periodicities, especially different temporal coverage, 
signal-to-noise ratio and time steps between the CSS and ZTF light curves, as well as the effects of red noises related to 
intrinsic AGN variability. Our analysis shows that red noises have strong influence on the different periodicities in 4C 50.43 
supporting our previous simulations. The results in this manuscript strongly indicate that it should be cautioned for applications 
of determined optical QPOs in BLAGN having strong intrinsic AGN variability. 
}


 \maketitle
	
\section{Introduction}
	
	Long-standing optical quasi-periodic oscillations (QPOs) observed in long-term light curves of broad line active 
galactic nuclei (BLAGN) have been proposed as potential indirect indicator of central sub-pc binary black hole systems 
(BBHs). Through applications of detected optical QPOs with periodicities ranging from hundreds to thousands of days, over 
200 sub-pc BBHs candidates have been reported. Large-scale time-domain surveys identified 111 potential candidates in 
\citet{gd15b} and confirmed 50 quasars exhibiting significant QPOs in \citet{cb16}. For studies of individual sources, 
PG 1302-102 exhibits periodicity of about 1800days \citep{gd15a}, PSO J334.2028+01.4075 shows periodicity of 542days 
\citep{lg15}, while SDSS J015910.05+010514.5 \citep{zb16} and SDSS J025214.67-002813.7 \citep{lc21} display periodicities 
of 1500days and 1607day, respectively. Compelling cases have also been found in Nearby AGN, such as Mrk 915 with 1150day 
QPOs \citep{ss20} and Mrk 231 with 1.2year QPOs \citep{ky20}. In addition, our recent works have reported several optical 
QPOs: 6.4year QPOs in SDSS J0752 \citep{zh22a}, 3.8year QPOs in SDSS J1321 \citep{zh22b}, 340day QPOs in SDSS J1609 
\citep{zh23a}, 1000day QPOs in SDSS J1257 \citep{zh23b}, and 550day QPOs in the quasar PG 1411+442 \citep{zh25b}.

	However, the authenticity of the QPOs remains controversial, although these studies provide important indirect 
evidence for the existence of sub-pc BBHs. Red noises related to intrinsic AGN variability may produce similar periodic 
features as first discussed in \citet{vu16} and in our recent works in \citet{zh23a, zh23b, lc25, zh25a}. Moreover, the 
effects of intrinsic AGN variability on the detected optical QPOs, especially whether it may cause the detected QPOs to 
significantly deviate from the real intrinsic QPOs, has not been discussed in depth. Our recent simulation study tackled 
this question in \citet{zh25b} that in 41.52\% of simulated light curves created in sub-pc BBHs, the detected QPOs deviated 
significantly from the input orbital period, often being less than two to four times the orbital period.

	 Although we \citep{zh25b} have proposed that AGN variability may affect the periodicity of QPOs, there are no 
relevant examples that can directly support this hypothesis. In this manuscript, we report on a distinctive source, 
4C 50.43, identified as a candidate of sub-pc BBH with an optical QPOs that have a reliable periodicity of about 1125days 
confirmed by \citet{gd15b} based on the Catalina Sky Survey (CSS). However, a significantly shorter periodicity of about 
513days is detected in its optical light curves collected from Zwicky Transient Facility (ZTF) \citep{ref8, ref9}. It 
is the first time to report such periodicity discrepancy of optical QPOs in different periods. The structure of this 
manuscript is as follows: Section 2 presents the analysis of optical QPOs in both CSS and ZTF light curves; Section 3 
discusses the possible explanations for the periodicity discrepancy; and Section 4 summarizes the main conclusions. 
Throughout the manuscript, we have adopted the cosmological parameters of $H_{0}$=70 km s$^{-1}$ Mpc$^{-1}$, 
$\Omega_{m}$=0.3, and $\Omega_{\Lambda}$=0.7.
	

\begin{figure*}
\centering\includegraphics[width = 8.5cm,height=4.75cm]{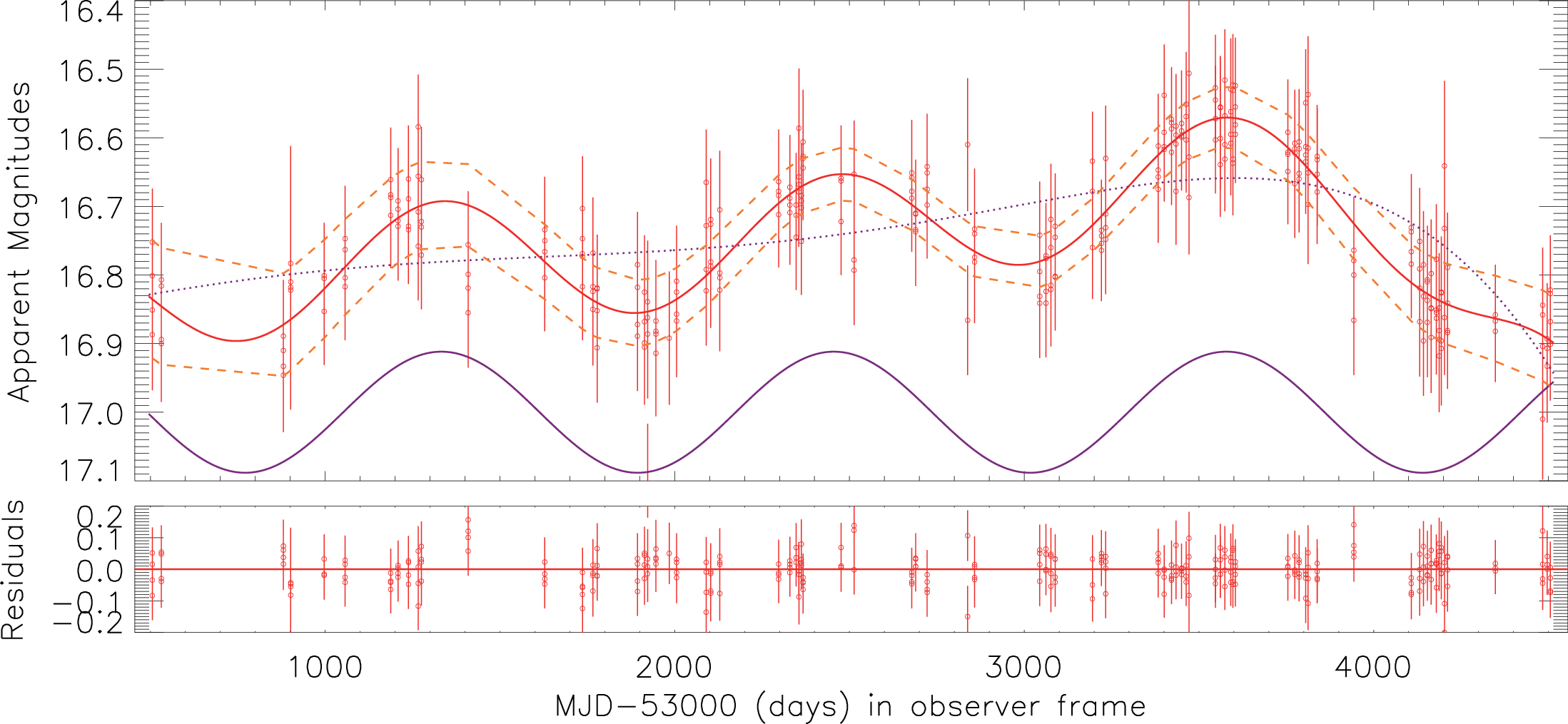}
\centering\includegraphics[width = 8.5cm,height=4.75cm]{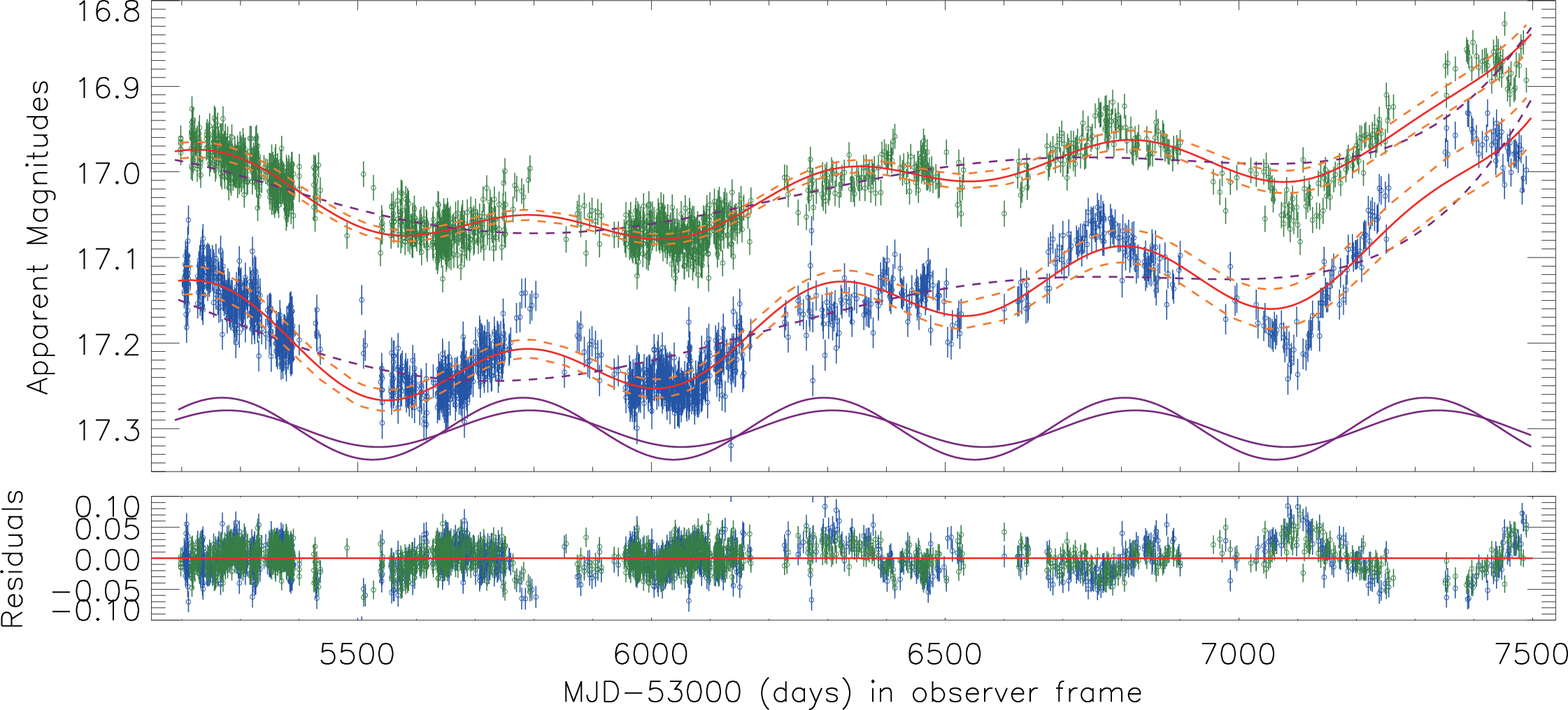}
\centering\includegraphics[width = 8.5cm,height=4.75cm]{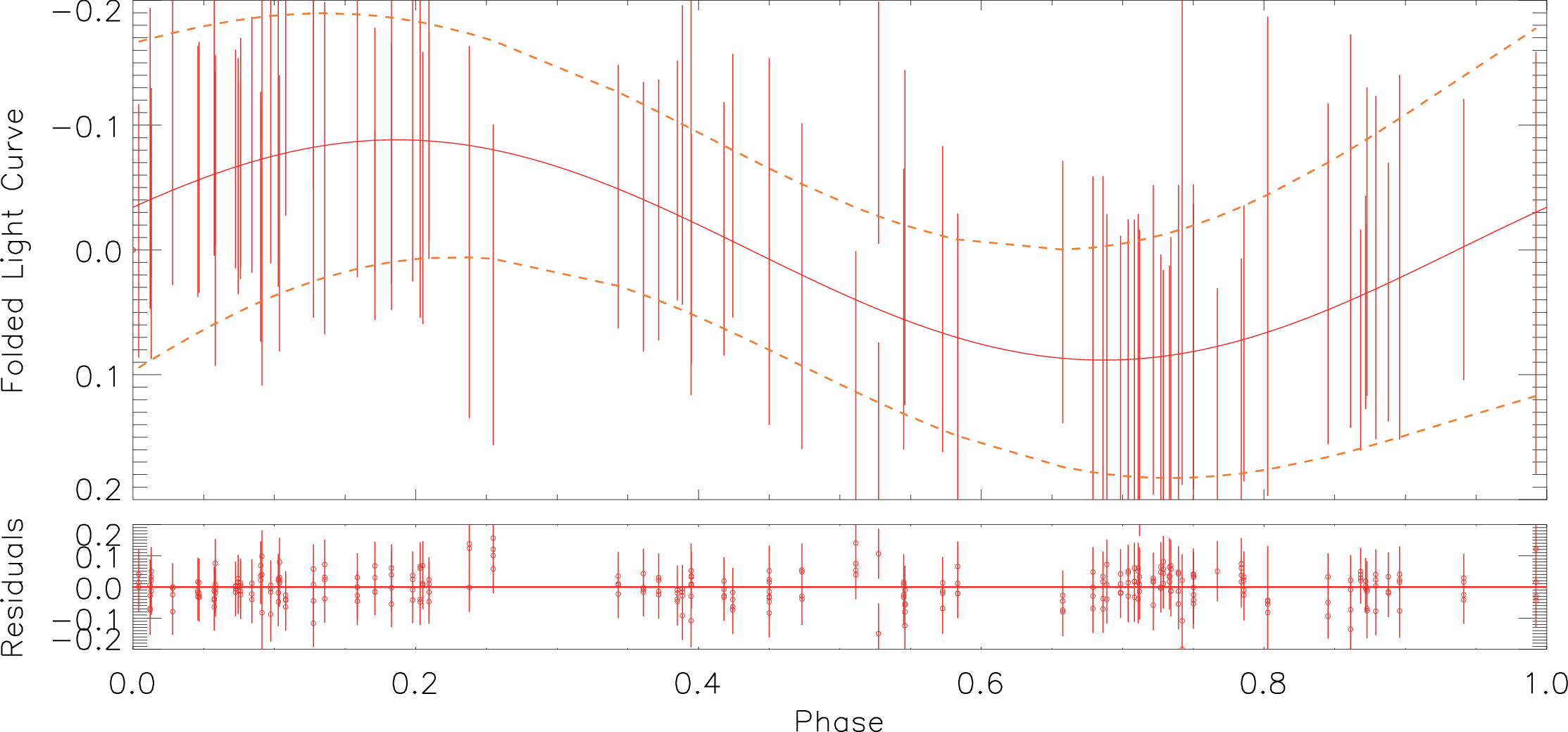}
\centering\includegraphics[width = 8.5cm,height=4.75cm]{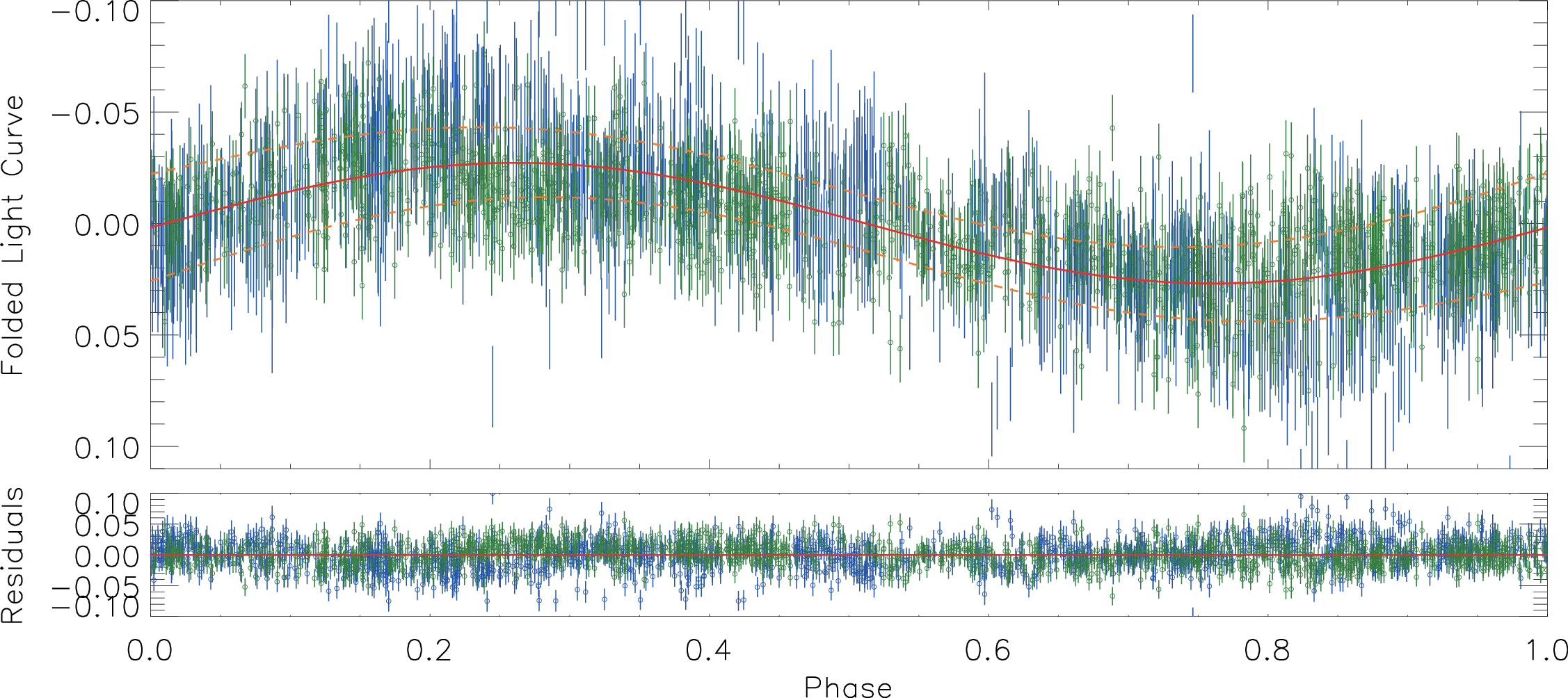}
\caption{Light curves fitting result of the CSS-V band (left, red symbols) and the ZTF g/r band (right, blue and green 
symbols). The first row presents the direct sinusoidal fitting results: in top panel, the solid red lines and orange 
dashed lines show the best-fitting results and corresponding 5$\sigma$ confidence bands from a sinusoidal function plus a 
fifth-degree polynomial component. The dashed purple lines represent the polynomial components, and the solid purple 
line represent the determined sinusoidal components. The second row show the phase-folded fitting results: in top panel, 
the solid red lines and dashed orange lines show the best-fitting results and the corresponding 5$\sigma$ confidence 
bands. The bottom panels of each row display the corresponding residuals (light curves minus the best fitting results), 
with solid red line showing residuals=0.}
\label{LMC}
\end{figure*}

\section{Optical QPOs in \obj}

	The CSS V-band light curve of \obj~ with MJD-53000 from 505 to 4505 (May 2005 to April 2016) and the ZTF g/r band 
light curves with MJD-53000 from 5204 to 7489 (March 2018 to June 2024) were collected and shown in top panels of 
Fig.~\ref{LMC}.  Here, the ZTF-i band light curve is not considered, as it contains relatively few data points. To improve 
the accuracy and reliability of optical QPOs, the direct fitting method, the phase-folding method, and the generalized 
Lomb-Scargle (GLS) method \citep{vt18} have been applied.

	The collected light curves are described by using a direct sinusoidal fitting approach that incorporates both 
long-term trend by a fifth-degree polynomial and periodic sinusoidal component. Here, the sinusoidal function is only 
used to model the periodic structure in the light curves, and does not involve discussions of physical origin of QPOs. 
Through the Levenberg-Marquardt least squares technique, the best fitting results can be determined and are shown in 
top panels of Fig.~\ref{LMC}, with \(\chi^2/\text{dof}\sim0.4\) and determined periodicity $1122\pm24$days in the CSS 
V-band light curve, and with \(\chi^2/\text{dof}\sim1.9\) and determined periodicity $512\pm16$days in the ZTF g/r-band 
light curves.

	Further, after subtracting the polynomial components, the light curves are phase-folded based on the determined 
periodicity. A sinusoidal function is then applied to describe the folded light curves, and the fitting results are shown 
in bottom panels of Fig.~\ref{LMC} with \(\chi^2/\text{dof}\sim0.4\) for the folded CSS-V band light curve and 
\(\chi^2/\text{dof}\sim2.1\) for the folded ZTF g/r band light curves.

\begin{figure}
\centering\includegraphics[width = 8cm,height=3.75cm]{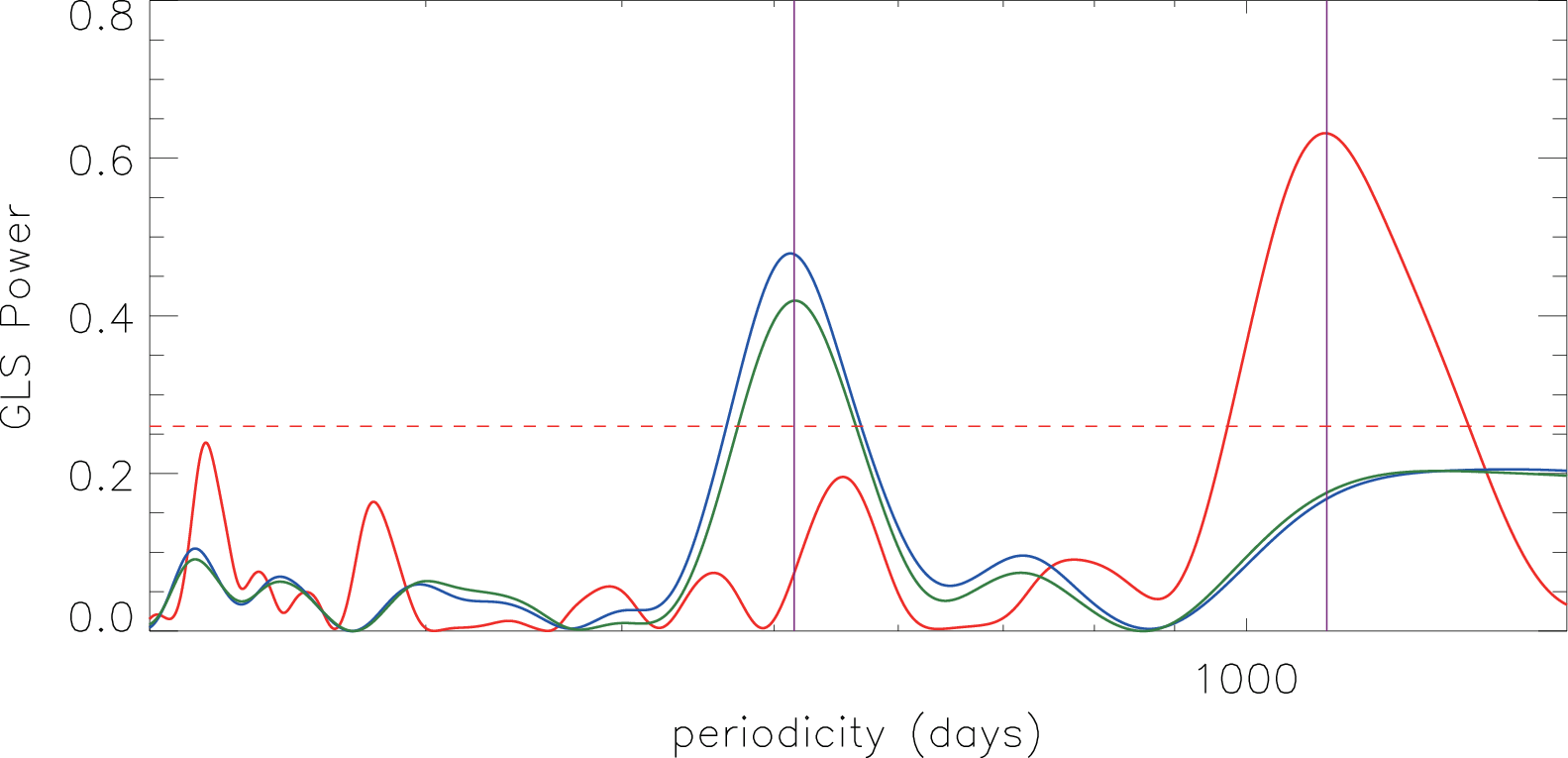}
\centering\includegraphics[width = 8cm,height=3.75cm]{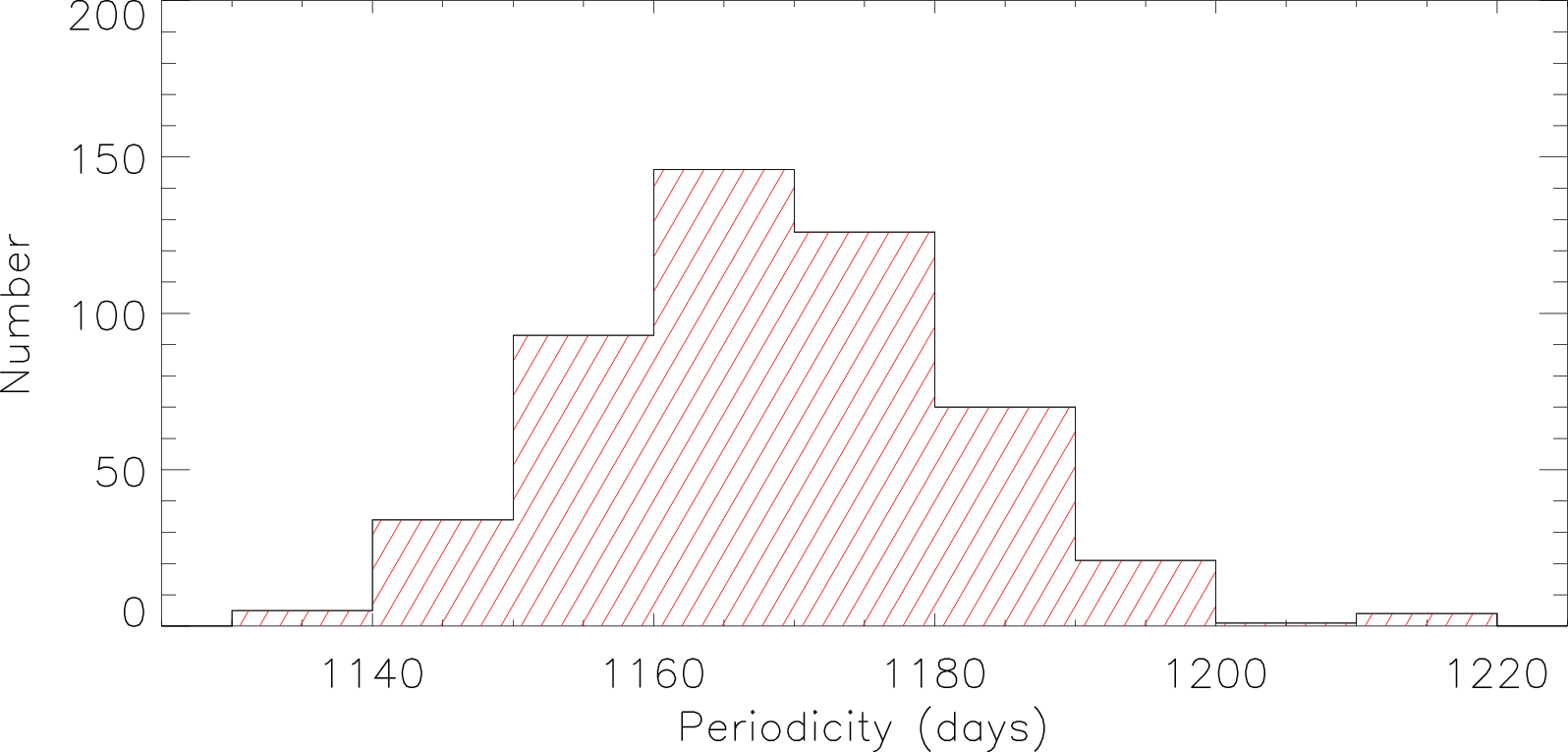}
\centering\includegraphics[width = 8cm,height=3.75cm]{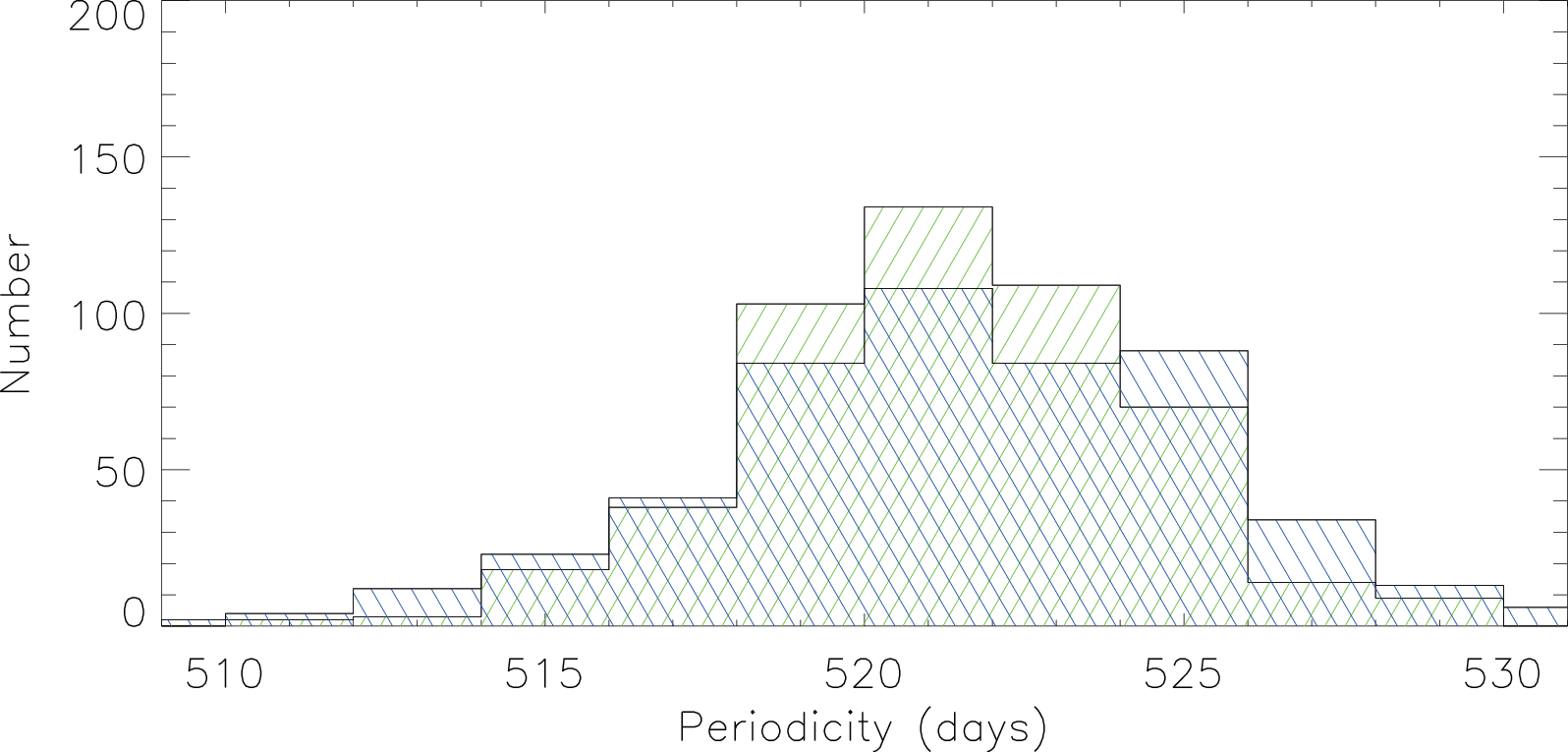}
\caption{Top panel shows the GLS periodograms of light curves. Solid lines in red, blue, and green show the corresponding 
results from the CSS V-band and ZTF r/g-band light curves. The horizontal dashed lines in red indicate the corresponding 
5\(\sigma\) significance level (FAP = 3e-7). Middle and bottom panels show distributions of the periodicity determined by 
the bootstrap method from the CSS V-band light curve and the ZTF g/r-band light curves.}
\label{gls}
\end{figure}

	In order to check the robustness of the QPOs, GLS method is employed. Top panel of Fig.~\ref{gls} presents the 
GLS powers of the light curves. A significant peak at approximately $1125$days is detected in the CSS-V band (solid red 
line), while the ZTF g/r bands (solid lines in blue and in green, respectively) exhibit prominent peaks around $515$days. 
All detected primary peaks exceed the 5$\sigma$ significance level (false alarm probability FAP=3e-7), as determined by 
the Bootstrap method.

	The uncertainty associated with the GLS determined periodicity is defined as half the full width at half maximum 
of the distribution of the peak periodicity obtained through bootstrap resampling within 1000 loops. Specifically, the 
original light curve is resampled 1000 times, with each resampled set randomly selecting more than half of the original 
data points. The bootstrap method determined distributions of the periodicities are shown in middle and bottom panels of 
Fig.~\ref{gls}.

	QPOs with periodicities of 1122$\pm$24days and 512$\pm$16days were detected in the CSS-V band and the ZTF g/r 
band light curves, using both the direct method and the phase-folded method. Similar periodicities of 1125$\pm$40days 
and 515$\pm$4days were also identified using the GLS method. The consistency of these results across different methods 
reinforces the reliability of the findings, collectively suggesting the presence of a significant QPOs with periodicity 
of around 1124$\pm$64days in the CSS-V band, consistent with that reported by \citet{gd15b}. While the ZTF g/r band show 
a periodicity of around 513$\pm$20days, it apparently differs from the one detected in the CSS-V band. Although the 
ratio of the two periodicities exhibits a near integer multiple of 2 relationship, the absence of simultaneous 
significant peaks around 1125days in the GLS periodogram(Fig.~\ref{gls}) of the ZTF g/r bands suggests a lack of harmonic 
coherence. 

	Before ending the section, the probability of optical QPOs in \obj~ related to intrinsic AGN variability (red 
noises) can be simply determined. The intrinsic AGN variability was simulated using the Continuous AutoRegressive (CAR) 
process \citep{kbs09}:
\begin{equation}
dLMC_t = \frac{-1}{\tau} LMC_t \, dt + \sigma_c \sqrt{dt} \epsilon(t) + bdt,
\end{equation}
where $\epsilon(t)$ denotes a white noise process with zero mean and unit variance. The mean magnitude $bdt$ was sampled 
from the typical range of 16 to 19 mag for low-redshift quasars. The relaxation timescale $\tau$ was randomly selected 
within the range of 100 to 1000 days, in line with typical quasar values reported in \cite{kbs09,m10}. The parameter 
$\frac{\tau\times \sigma_{*}^2}{2}$, which corresponds to the variances of simulated light curves, was randomly sampled 
within the typical range of variance values in the ZTF light curves of SDSS quasars, from 0.001 to 0.32. The uncertainties 
$LMC_{err}$ of the simulated light curves $LMC$ are determined based on the relative uncertainties in the observed light 
curves, $\frac{LMC_{err}} {LMC}=\frac{LMC_{obs,err}}{LMC_{obs}}$. The time information $t=[t_{css}, t_{ztf}]$ is the 
combined time information of CSS V-band and ZTF g-band light curves of \obj. Then, among 1e6 simulated light curves, there 
are 113 light curves collected according to the following criteria. The GLS determined periodicity in the simulated light 
curve with $t=t_{css}$ ($t=t_{ztf}$) should be larger than 1124-128days (513-80days) and smaller than 1124+128days 
(513+80days) and have the corresponding peak value in the GLS powers higher than 0.5 (0.4) (the peak values larger than 0.6 
and 0.5 in CSS and ZTF in \obj). Therefore, the probability should be 99.99\% (1-113/1e6), the corresponding confidence level 
higher than 4$\sigma$, that the detected optical QPOs in \obj~ are not from intrinsic AGN variability.

\begin{figure*}
\centering\includegraphics[width = 8cm,height=3.75cm]{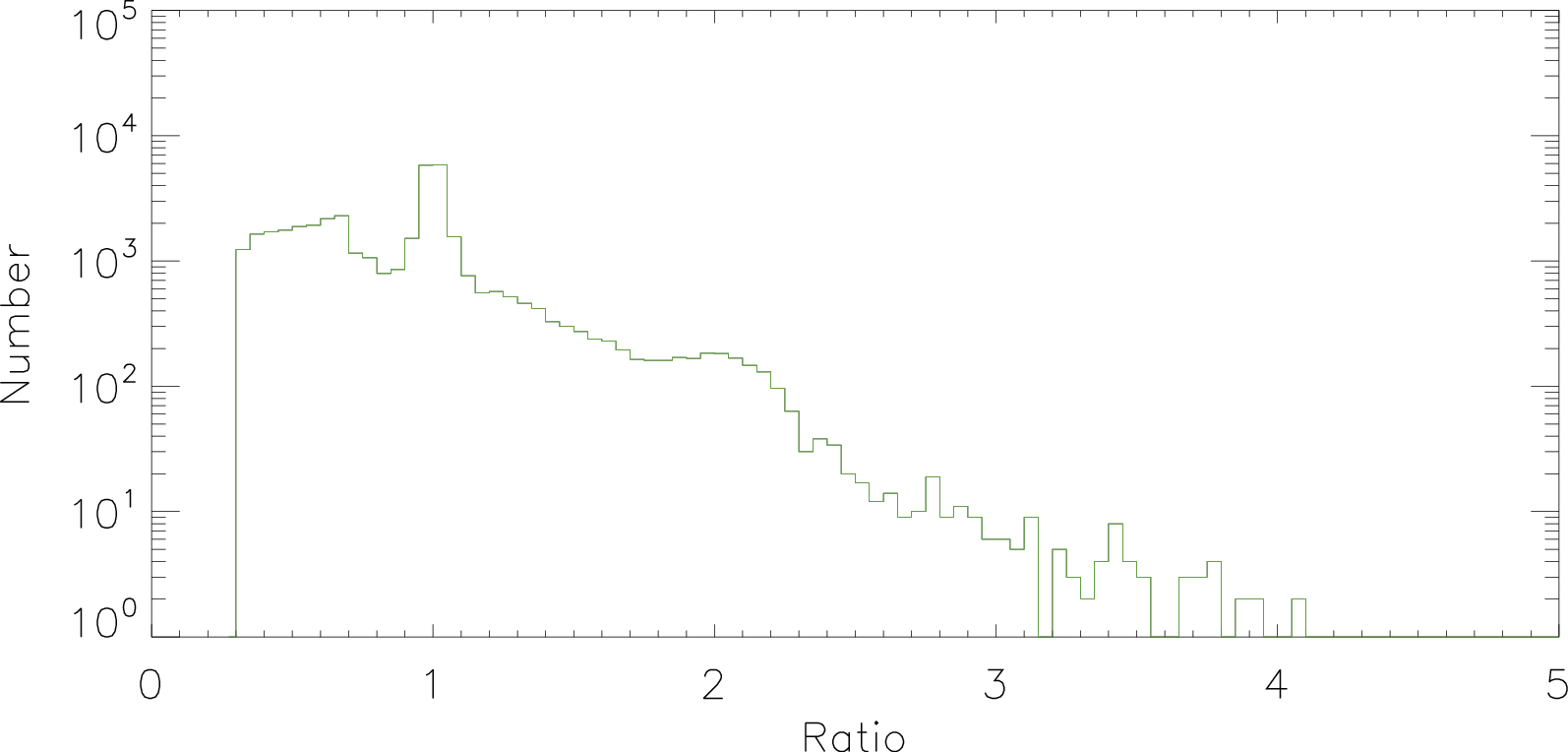}
\centering\includegraphics[width = 8cm,height=3.75cm]{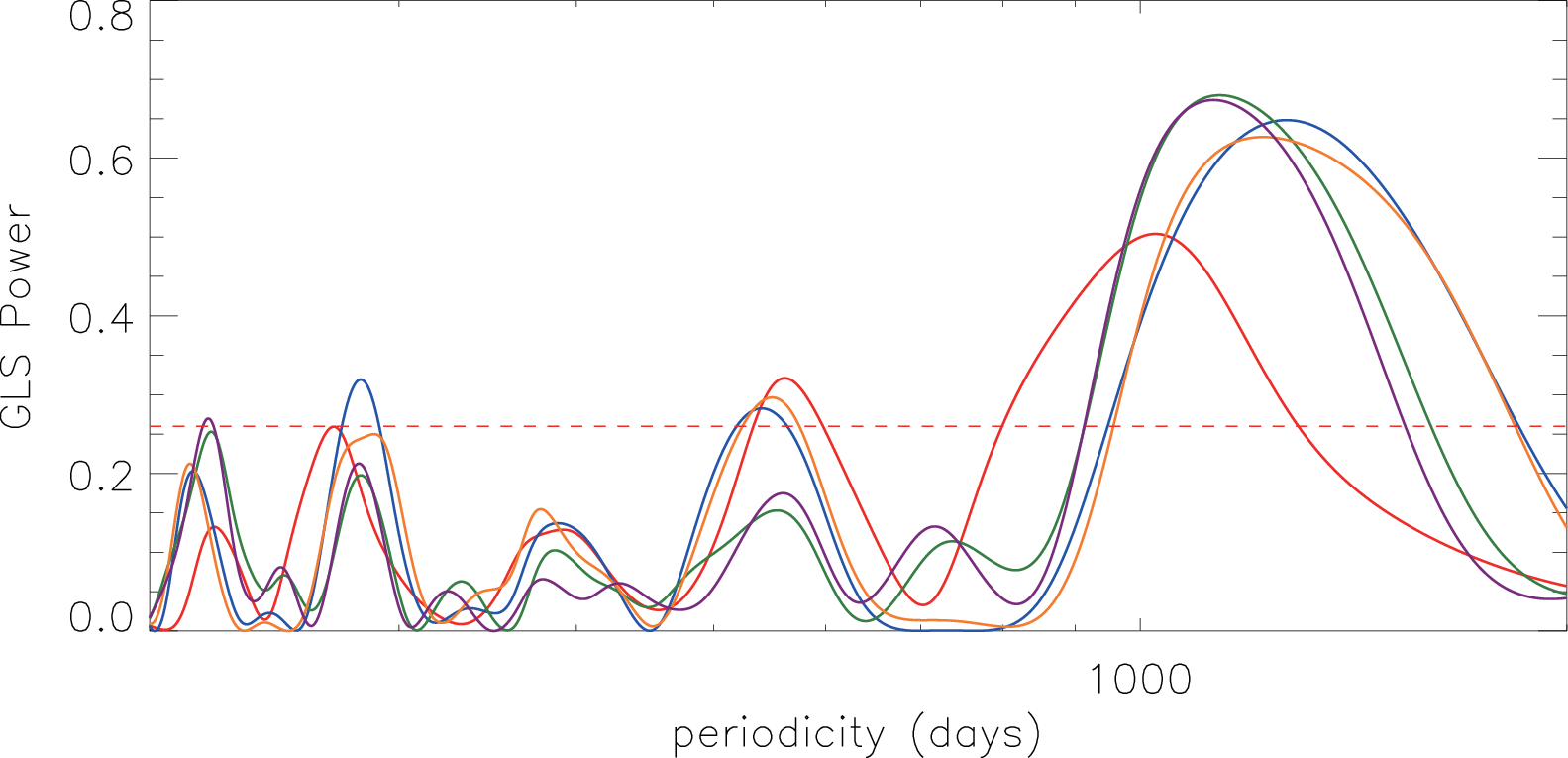}
\centering\includegraphics[width = 8cm,height=3.75cm]{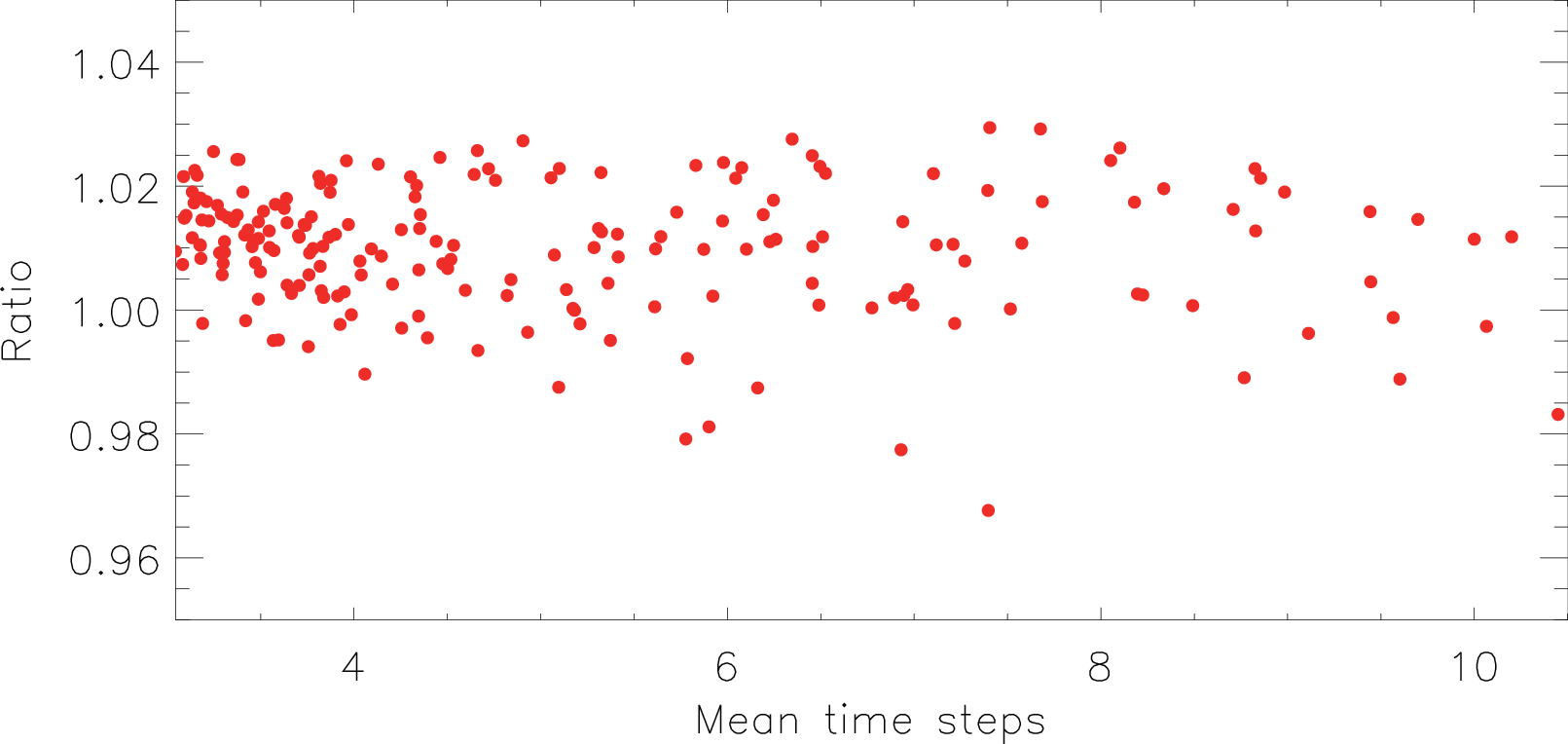}
\centering\includegraphics[width = 8cm,height=3.75cm]{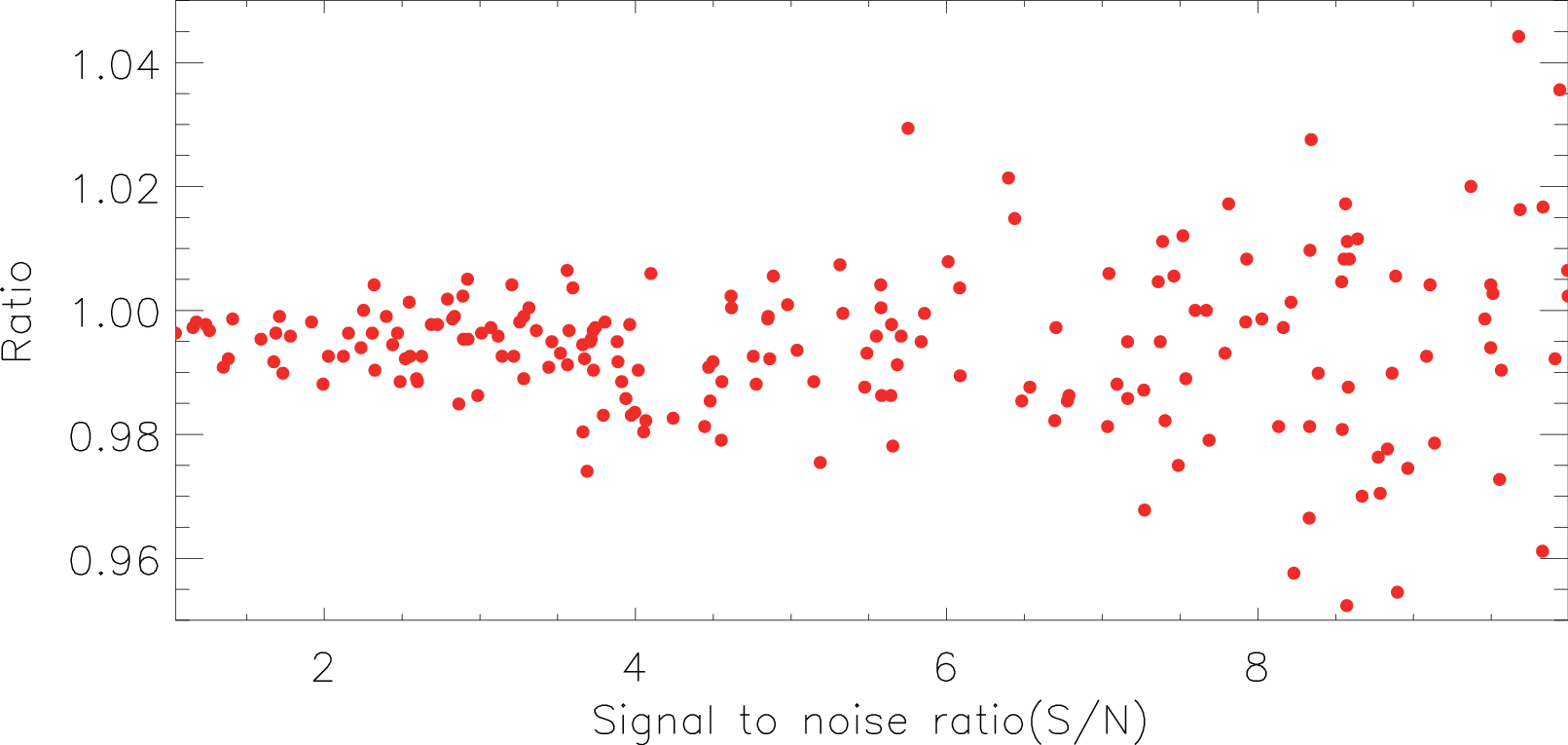}
\caption{Upper left panel shows distribution of the ratio of the periodicity $T$ to the measured periodicity detected 
by GLS method applied in the simulated light curves including sine component and intrinsic AGN variability. Upper right 
panel shows the GLS periodograms of five light curves randomly selected from the CSS V-band light curves, with time spans 
2236days, 2475days, 2217days, 2208days, and 2466days (in orange, green, red, blue, and purple, respectively), similar 
to that of the ZTF g band. Lower left and right panels show periodicity ratios between each subset with varying time 
steps (lower left) and varying signal-to-noise ratios (lower right) and the original ZTF g band light curve. }
\label{corr}
\end{figure*}

\section{Potential causes of QPOs detection inconsistencies between CSS-V and ZTF g/r band}

	 Given the significant wavelength overlap between the CSS-V and ZTF g/r bands, both probe similar regions of the 
accretion disk. The periodicity discrepancy is therefore unlikely to reflect radial dependence; instead, it is more likely 
attributable to red noise, as well as variations in S/N, temporal coverage, and time steps between the CSS and ZTF light curves.

	As detailed discussions in our recent work \citet{zh25a}, the effects of intrinsic AGN variability should be 
accepted to explain the periodicity discrepancy of optical QPOs in different wavelength bands in different periods in \obj. 
Here, an additional oversimplified method is applied to re-check the effects of red noises related to intrinsic AGN variability. 
The intrinsic AGN variability (red noises) was similarly simulated by the CAR process, but with time information to be 
$t_{ztf}$ of the ZTF g-band light curve of \obj. And then the intrinsic QPOs was introduced into the red-noise-dominated 
light curves in the form of a sinusoidal function $S(t) = A \times \sin\frac{2\pi t}{T}$ with amplitude A randomly selected 
from the set [0.1, 0.2, 0.3, 0.4, 0.5, 0.6, 0.7, 0.8, 0.9, 1, 2, 3, 4, 5, 6, 7, 8, 9] multiplied by the variance of the red 
noise. The periodicity T is randomly selected from 300days to 700days. The simulated light curves including QPOs and intrinsic 
AGN variability can be created by $S(t) + LMC_t$.

	Based on the above method, a total of 50000 light curves were generated with the same information $t$ as the ZTF g 
band light curve of \obj~ (including the observation timestamps and temporal sampling), due to its high quality. The corresponding 
periodicities of these simulated light curves were detected using the GLS method. Upper left panel of Fig.~\ref{corr} presents 
the distribution of the ratio of the intrinsic periodicity $T$ to the measured periodicity $T_{o}$ (with a significance level 
exceeding $5\sigma$) derived from the simulation light curves using the GLS method. The results indicate that the majority of 
the ratios are inconsistent with 1, leading the influence of red noises on the intrinsic QPOs to be apparent. Meanwhile, among 
the 50000 simulated light curves, 37152 exhibited a periodicity ratio $T/T_{o} \leq 1.5 $. Meanwhile 3042 showed substantial 
deviations, with some ratios reaching up to 4, suggesting that red noises significantly affects the accuracy of periodicity 
measurements. The remaining light curves were excluded from the statistical analysis due to the lack of statistically significant 
periodicity. The results strongly indicates that effects of red noises can be applied to naturally explain the periodicity 
discrepancy in \obj.

	Considering the temporal coverage of the ZTF light curves may not be sufficient to detect the 1124days periodicity 
with statistical significance. Five light curves with time spans similar to that of the ZTF g band were randomly selected 
based on the existing CSS-V band light curves. The time spans of these light curves are 2236days, 2475days, 2217days, 2208days, 
and 2466days, with the corresponding data points being 152, 220, 138, 154, and 226, and sampling frequencies of 0.068, 0.089, 
0.062, 0.07,and 0.09 respectively. GLS analysis was then performed on each of these light curves, with the results shown in 
upper right panel of Fig.~\ref{corr}.

 	The analysis reveals that QPOs in all five light curves are concentrated around 1124days, indicating that such time 
spans are sufficient for identifying this periodicity. Furthermore, higher sampling frequencies yield more significant detection 
of the 1124days QPO, while reducing the significance of the 513day periodicity. Given that the ZTF light curves possess 
comparable temporal coverage and sampling frequencies, the 1124days periodicity should be detectable if it were truly present. 
However, the ZTF light curves exhibit a dominant peak at 513days with no significant detection at 1124days. Therefore, the 
absence of the 1124days periodicity in the ZTF light curves cannot be attributed to insufficient temporal coverage; rather, 
this periodicity likely does not exist in the ZTF.

	Due to the smaller time steps and higher signal-to-noise (S/N) ratios of ZTF light curves than CSS light curves, it 
is necessary to check whether quality of light curves can be applied to explain the detected periodicity discrepancy in \obj, 
especially effects of time steps and signal-to-noise ratios.

	For effects of the time steps, 200 subsets containing 1/6 to 3/4 of the original data points were randomly selected 
from the ZTF g-band light curve of \obj. For effects of the S/N, the original photometric uncertainties were scaled by randomly 
selected factors between 1 and 10 and added to the ZTF g-band light curve of \obj, with this process repeated 200 times. 
The periodicities of both two sets of samples was determined by GLS method, and the resulting periodicities were compared 
to that of the original light curve in terms of their ratio. As shown in the bottom panels of Fig.~\ref{corr}, the results 
reveal that variations in time steps have only a limited impact on QPOs detection, with the periodicity remaining largely 
consistent with that of the original ZTF g-band light curve of \obj. Meanwhile, there were also few effects of S/Ns on detected 
periodicities, leading to only around 5\% difference between intrinsic periodicities and determined values. Therefore, 
quality of light curves has few effects on the detected periodicity discrepancy in \obj.

\section{CONCLUSIONS}
	We report the quasar \obj~ exhibiting a significant discrepancy in periodicity of optical QPOs that is unlikely to 
represent a harmonic relationship: 1124days in the CSS Vband and 513days in the ZTF g/r bands, as revealed by different 
methods. Through simulations by CAR process, confidence level higher than 4$\sigma$ can be confirmed that the detected 
optical QPOs in \obj~ are not related to intrinsic AGN variability. Several possible mechanisms were considered to explain 
this difference, such as different  temporal coverage and quality of light curves, as well as the effects of red noises. In 
conclusion, the periodicity difference is more likely caused by intrinsic AGN variability. At current stage, although the 
unique periodicity discrepancy only reported in \obj, the results in the manuscript strongly indicate it should be cautioned 
for applications of optical QPOs in BLAGN especially having strong intrinsic variability leading to detected periodicity 
very different from intrinsic values.

\begin{acknowledgements}
The authors gratefully acknowledge the anonymous referee for giving us constructive comments to greatly improve the paper. 
Zhang gratefully acknowledges the kind grant support from NSFC-12373014 and 12173020 and the support from Guangxi Talent 
Programme (Highland of Innovation Talents). This paper has made use of the data from ZTF and CSS, and the MPFIT package, 
and the NASA/IPAC Extragalactic Database (NED).
\end{acknowledgements}

\label{lastpage}
\end{document}